\documentstyle[aasms4]{article}

\begin{document}
\title{EVOLUTION AND NUCLEOSYNTHESIS OF ZERO METAL INTERMEDIATE MASS STARS}
  
\author {Alessandro Chieffi}
\affil{Istituto di Astrofisica Spaziale (CNR), via Fosso del
 Cavaliere, I-00133 Roma, Italy\\achieffi@ias.rm.cnr.it}
\author {Inma Dom\'{\i}nguez}
\affil{Dpto. de F\'{\i}sica Te\'orica y del Cosmos, Universidad de Granada, 18071 Granada, Spain\\ inma@ugr.es}
\author {Marco Limongi}
\affil{Osservatorio Astronomico di Roma,
Via Frascati 33, I-00040 Monteporzio Catone (Roma), Italy\\marco@nemo.mporzio.astro.it}
\and
\author {Oscar Straniero}
\affil{Osservatorio Astronomico di Collurania,
I-64100 Teramo, Italy\\straniero@astrte.te.astro.it}

\begin{abstract}

New stellar models with mass ranging between 4 and 8 M$_{\odot}$, Z=0 and 
Y=0.23 are presented. The models have been evolved from the pre Main Sequence up to the Asymptotic Giant 
Branch (AGB). At variance with previous claims, we find that these updated 
 stellar models {\it do} experience thermal pulses in the AGB phase. In particular we
 show that:
 
a) in models with mass larger than 6 M$_{\odot}$, the second dredge up is able to 
raise the CNO abundance  
in the envelope enough to allow a {\it normal} 
AGB evolution, in the sense that the thermal pulses and the third dredge up 
settle on;

b) in models of lower mass, the efficiency of the CNO cycle in the H-burning shell 
is controlled by the carbon
produced locally via the 3$\alpha$ reactions. Nevertheless  
the He-burning shell becomes thermally unstable after the early AGB.
The expansion of the overlying layers induced by these weak He-shell flashes is not sufficient
by itself to allow a deep penetration of the convective envelope. However, 
immediately after that, the maximum luminosity of the He flash is
attained and 
 a convective shell systematically forms at the base of the H-rich envelope. 
The innermost part of this convective shell probably overlaps the underlying C-rich region 
left by the inter-shell convection during the thermal pulse, so
that fresh carbon is dredged up in a {\it hot} H-rich environment and a H flash occurs.
This flash favours the expansion of the outermost layers already started by the weak
 thermal pulse
and a deeper penetration of the convective envelope takes place.
Then,
the carbon abundance in the envelope rises to a level high enough that the
 further evolution of these models
closely resembles that of more metal rich AGB stars.

These stars provide an important source of primary carbon and nitrogen, so 
 a major revision of the chemical evolution in 
the early Galaxy is required.
We suggest that the chemical imprint of these Pop III stars could be found in 
the old and  metal poor components of the Milky Way.

\end{abstract}

\section{Introduction}

Evidence for an expanding Universe coupled to the observation of the fossil black body radiation
leads to the natural conclusion that the primordial Universe was hot and dense and that there was an epoch
during which nuclear reactions among neutrons and protons took place. A few 
hours after the Big Bang the primordial composition of the Universe was defined. 
The standard homogeneous Big Bang nucleosynthesis (as first computed by Fermi \& Turkevich, for a recent
analysis see Walker et al. 1991) predicts
that the material which emerged from this epoch was mainly made up of $^1$H, $^4$He and a small quantity 
of other light elements (D, $^3$He and $^7$Li). The total mass fraction of heavier elements was lower than
$10^{-10}$, so that the first generation of stars, the so called population III,  
was formed from a gas essentially lacking of metals.
Such a peculiarity certainly influenced both the fragmentation of 
the primordial gas clouds and the evolutionary properties of this first stellar generation.
In fact, in these extreme conditions, the cooling able to reduce 
the Jeans mass down to the stellar values was
provided by the molecular hydrogen rather than by dust or heavy molecules and the progressive fragmentation was
halted only when the gas became opaque due to the H$_2$ line absorption (Carlberg 1981; 
Silk 1983; Lepp \& Shull 1983; Palla, Salpeter \& Stahler 1983; 
Yoshii \&  Saio 1986; Shapiro \& Kang 1987; de Ara\'ujo \& Opher 1989; 
Uehara, Susa \& Nishi 1996; Haiman, Thoul \& Loeb 1996; Omukai et al. 1998).

The initial mass function (IMF) emerging from this atypical star formation process  
has been the subject of a number of papers.
  Several groups have found  
that intermediate mass stars (IMS) could result from the fragmentation of primordial
gas clouds. Yoshii and Saio (1986) found that the peak of the IMF
for population III stars ranges between 4 and 8 $M_{\odot}$ while 
Uheara et al. (1996) derived that the minimum stellar mass is  
of the order of the Chandrasekhar mass (i.e. $\sim$1.4 $M_{\odot}$).  
Nakamura \& Umemura (1999) concluded that the typical stellar mass
 is around 3M$_\odot$; this value may further increase  
by accretion of the environmental gas up to 16 M$_\odot$. 
On the other hand, 3D simulations (Abel et al. 1997; Abel, Bryan \& Norman 2000; 
Bromm, Coppi \& Larson 1999) obtained a Jean mass of the order of 10$^2$-10$^3$ $M_{\odot}$;  
  their calculations, however, do not have enough resolution to explore further 
 fragmentation into smaller masses. 
 Recently, Nakamura \&    
 Umemura (2000) have performed 2D simulations and have found that the primordial IMF is likely to be bimodal, the first peak around 
 2 $M_{\odot}$ and the second peak between  
  10 and 10$^2$ $M_{\odot}$, consistent with the previous cited works.
Wasserburg \& Qian (2000a,  2000b), by comparing iron and r-element abundances in very metal poor stars, have concluded that the first generation of galactic stars was essentially composed  of massive objects (M$\ge$100 M$_{\odot}$) capable of producing a prompt iron synthesis. Following their model, the average galactic metallicity would increase up to [Fe/H]$\sim$-3 in a few $10^6$ yr. In such a case even low mass stars could form after a very short time,
but their iron content would be that of population II stars.
So the actual IMF of population III stars is still an open question.

However, since the chemical composition of the matter ejected by a star largely
depends on its mass, the study of the surface chemical composition of the most metal poor stars
could help to shed light on the kind of stars which first populated our galaxy (and probably the Universe).
Note, by the way, that the lifetime of the more massive IMS is short enough so that they could have really polluted the
interstellar medium during the very early phase of dynamical collapse of our galaxy (masses larger than, say,
$\simeq4$  M$_{\odot}$ live less than $\simeq10^8$ yr).

Great observational efforts have been made to identify 
very metal deficient stellar populations. The first attempt (Bond, 1981) failed to find stars with metallicity 
[Fe/H]$<$-3. Later on, a red giant with [Fe/H]$\sim$-4.5 
was identified by Bessel and Norris (1984) and since then the number of known low metallicity stars 
 has increased somewhat (see e.g.  Beers, Preston \& Shectman 1992; Ryan, Norris \& Bessel 1991) though
only 6 of them have [Fe/H]$\le$-3.5. Some 
authors suggest that 4 of these 6 stars have a metallicity
[Fe/H]$\sim$-4 (see Ryan, Norris \& Beers 1996, 1999; Sneden et al. 1994; Primas et al. 1994;  
Ryan et al. 1991; Carney \& Peterson 1991; Molaro \& Castelli
1990; Molaro \& Bonifacio 1990). To date, however, no stars having the primordial
chemical composition
(i.e. Z${\sim}10^{-10}$ or [Fe/H]$\sim$-8.3) have been found.
Such a lack of "observable zero metal" stars actually supports the theoretical
predictions that the primordial IMF did not favour the formation of low mass, long lived stars,
but it does not rule out the primordial intermediate mass stars (P-IMS).

Though there is definite evidence that very metal poor stars (and also just metal poor stars)
formed by matter strongly polluted at least by the ejecta of massive stars (consider, for 
example, the overabundance
of the $\alpha$ elements with respect to iron) there is some hint that a P-IMS generation formed:
as an example, let us mention that the galactic nitrogen behaves as a primary element, i.e.
it is solar scaled (although with an important dispersion around [N/Fe]$\sim$0.0) in the lower metallicity stars observed in the Milky Way
(Laird 1985; Carbon et al. 1987). While the nucleosynthesis occurring in massive stars, exploding
as type II supernovae, cannot account for this primary nitrogen component
(see, for example, Timmes, Woosley \& Weaver 1995), the presence of a generation of P-IMS could robustly
contribute to the production of primary nitrogen.

It goes without saying that the inclusion or otherwise of the ejecta of a P-IMS generation in the general
chemical evolution of a galaxy also alters drastically, the interpretation of the
heavy element abundances measured in the very low metallicity galactic stars 
 and the intergalactic medium (IGM) at the
high-redshifts (from z$\sim$2 to 4; Cowie et al. 1995;
Tytler et al. 1995; Songaila \&  Cowie 1996; Pettini et al. 1997; Ellison et al. 2000).

For all these reasons we
decided to revise the evolutionary properties and the related nucleosynthesis of these primordial stars
in a wide range of stellar masses. In this way, we hope to
identify the possible footprints left by the primordial stellar population in the galactic chemistry.

The necessity to calculate new P-IMS models derives from the fact that most  
of the available theoretical studies of this kind of stars stop 
at the end of the He-burning (Ezer 1961, 1972; Ezer \& Cameron 1971; Eryurt-Ezer 1981; Castellani, Chieffi \&
Tornamb\`e 1983; Tornamb\`e \& Chieffi 1986; Cassisi \& Castellani 1993).
As is well known, the most important contribution of the IMS
to the galactic chemical evolution comes from the nucleosynthesis occurring during the thermally pulsing
phase of the AGB.
These stars, in fact, produce in their interior carbon, nitrogen and neutron rich isotopes which 
 can 
 enrich the interstellar medium by the combined efforts of the dredge up and the mass loss.
In spite of their potential importance in nucleosynthesis and chemical evolution, to date 
just one paper has addressed the computation of AGB models of this generation of P-IMS
(Chieffi \& Tornamb\'e 1984), the main finding (based on the evolution of a
5 $M_{\odot}$) being that these stars do not experience the thermally pulsing phase (see below).
In a companion paper Fujimoto et al. (1984) addressed the general behaviour
 of the helium shell flashes in zero metal AGB stars by means of an analytical model: 
their conclusion was that
the thermally pulsing phase takes place in stars less massive than  4 $M_{\odot}$.

The scenario which emerged from those computations was that the overall impact of a
P-IMS generation on the chemical evolution of the light elements (mainly C and N) in the galaxy 
was marginal.
On the other hand, the lack of thermal pulses coupled to a reduced mass loss rate (due to the fact that 
these stars are significantly more compact and dimmer than the present ones) led the authors to suggest 
that these stars were more likely to increase the    
degenerate C-O core mass up to 1.4 $M_{\odot}$ and hence to explode as type I ${1\over 2}$ 
supernovae. The result would  be a strong pollution of the interstellar medium of material formed by elements
produced by complete explosive Si burning (Fe, Co and Ni), incomplete explosive Si burning (Mn, Cr and V) and
explosive O burning (Si, S, Ar, Ca and K).

In this paper and in a companion one (Limongi, Straniero \& Chieffi 2001) we 
present a full set of evolutionary models of  zero metal stars 
ranging in mass between  4 and 25 $M{_\odot}$, from the pre main sequence up to the AGB or up 
to the iron core collapse. In particular this paper illustrates our results for models having
$M{\le}M_{up}$ (namely the smallest mass which ignites carbon off-centre in the degenerate core).
 
\section{The models}
 
The evolution of five P-IMS models, with mass 4, 5, 6, 7 and 8 M$_{\odot}$, Z=0 and Y=0.23,  
has been computed starting from the pre-main sequence up to either an advanced phase on the AGB
or up to the carbon ignition for the more massive model. 
These computations have been performed by 
means of the latest version of the FRANEC code, release 4.8 (see Chieffi, Limongi \& Straniero 1998). 
Detailed references for the input physics adopted have been reported 
by Straniero, Chieffi \& Limongi (1997).
The main characteristic of this code is that the set of equations
used to describe both the chemical evolution (due to the nuclear burning)
and the physical structure are fully coupled and solved simultaneously.
This coupling is required to follow the final phases of the evolution of massive stars, from
the oxygen burning onwards. In order to produce a homogeneous set of models, we have 
adopted the same computational scheme for intermediate mass stars.
The nuclear network used in the present computations includes 48 isotopes (269 reactions)
for the H-burning, 
and 34 isotopes (147 reactions) for the He-burning (see Fig. 1). In order to
identify the lowest stellar mass for which a degenerate carbon ignition occurs
($M_{up}$), a reduced set of nuclear species and related reactions
has been added for the carbon burning, namely 9 isotopes and 8 reactions.
A revised version of the time dependent mixing scheme first introduced by Sparks \& Endal (1980)
has been adopted. In particular the mass fraction ($X_j$) of a certain isotope 
at the mesh-point $j$, inside a convective region having total mass $M_{conv}$,
is given by:
\begin{equation}
X_j=X_j^o+\frac{1}{M_{conv}}{\sum}_{k}(X^o_j-X^o_k)f_{j,k}{\Delta}M_k
\end{equation}
where the summation is extended over the whole convective region and the
superscript $^o$ refers to unmixed abundances. ${\Delta}M_k$ is the mass of the mesh-point $k$,
while the damping factor $f$ is:
\begin{equation}
f_{j,k}=\frac{{\Delta}t}{{\tau}_{j,k}}    
\end{equation}
if ${\Delta}t<{\tau}_{j,k}$,  or
\begin{equation}
f_{j,k}=1                  
\end{equation}  
if ${\Delta}t\ge{\tau}_{j,k}$.
Here ${\Delta}t$ is the time step and $\tau_{j,k}$ is the mixing turnover time
between the mesh-points j and k, namely: 
\begin{equation}
\tau_{j,k}=\int_{r(j)}^{r(k)} \frac{dr}{v(r)}={\sum}_{i=j,k}\frac{{\Delta}r_i}{v_i}   
\end{equation}
The mixing velocity ($v_i$) is computed according to the mixing length theory (Cox \& Giuli 1968) and
${\Delta}r_i$ is the length of the mesh-point i.
This algorithm allows us to account for the partial mixing that occurs when the time step 
is reduced to or below the mixing timescale. This condition is sometimes fullfilled 
in the computation of advanced evolutionary phases. 
For example,  
thermally pulsing models require time steps of a few days 
(in some cases just a few hours)
which may be  
comparable to or lower than the mixing timescale defined in equation (4). 
The borders of the convective regions are identified by means of the classical Schwarzschild
criterion.
Note that we have used, for the H-rich material, opacity tables computed by
assuming $Z=10^{-4}$ and various ratios for the H and He abundances
 (Alexander \& Fergusson, 1994; Iglesias, Rogers \& Wilson 1992).
These tables allow a suitable description of the radiation transport 
in the stellar envelope,  as long as the
metallicity does not exceed $Z=10^{-4}$. For this reason, our evolutionary sequences
become unreliable when the metallicity in the envelope, as modified by the third dredge up,
increases over this value. 

The computation of stellar models is usually based on semi-empirical
(or phenomenological)
descriptions of hydrodynamic phenomena occurring in real stars,  
namely convection and mass loss. The widely used {\it mixing length theory} of convection 
is a well known example of this phenomenological approach to stellar hydrodynamics.
This procedure does not provide, in general, a
satisfactory physical description of these complex phenomena. Nevertheless, once 
the stellar models are properly calibrated, it may allow us  
to obtain adequate predictions for the stellar properties. 
This may be done by {\it tuning} the models (i.e. by changing the parameters left free
in the phenomenological theory) until they reproduce
 certain selected and well measured observable quantities.
Unfortunately, due to the lack of an observational counterpart, such a procedure cannot 
be followed
for Z=0 models, so that
the question of the calibration becomes particularly critical for population III stars.
A possible alternative approach consists of adopting the values of the tuned free parameters 
obtained for larger metallicities. However, owing to
the particular structure developed by metal deficient stars, 
an extrapolation of the calibrated parameters down 
to Z=0 may induce a substantial error. 
Thus, in order to 
estimate the uncertainties implicit in the above description of the stellar hydrodynamics, 
one should explore a range of values for these free parameters that is as wide as possible, but
this would imply the computation of a great number of stellar models. 
 A general discussion of 
this very important problem is beyond the goals of the present paper. 
In the following we will focus
our attention on those phenomena that particularly affect the estimated nucleosynthesis 
of the P-IMS stars. 

Mass loss has a negligible effect on intermediate mass stars up to 
the onset of the AGB. Strong mass loss rates (between $10^{-8}$ and 
$10^{-4}$ $M_{\odot}$/yr) are indeed observed in AGB stars.
The duration of this phase, the mass of the
resulting compact stellar remnant and the chemical yields are then substantially affected
by mass loss. Groenewegen \& De Jong (1993, 1994a, 1994b), on the basis of synthetic AGB models,
found that a Reimers (1975) mass loss rate with a multiplicative factor $\eta \sim 4$ 
can reproduce the observed high-luminosity tail of the carbon stars 
luminosity function in our Galaxy and in the 
Magellanic Clouds as well as the observed abundances in LMC planetary nebulae.
Dom\'\i nguez et al. (1999) have shown that with such a mass loss rate the final masses 
of intermediate mass stellar models are compatible with the initial-final
mass relation (Weidemann 1987; Herwig 1995). More recently other formulas 
have been proposed (Vassiliadis \& Wood 1993; Bl\"ocker 1995) that predict 
 a huge mass loss rate for AGB stars (see, for a recent revision of the initial-to-final mass relation, Weidemann 2000).   
We do not know if a similar strong mass loss rate also 
characterizes population III AGB stars.  It has been suggested (see, for example,  Vassiliadis and Wood) that
the increase of the mass loss rate during the AGB may be caused by the onset of large-amplitude 
radial pulsation. Whether P-IMS stars also suffered this kind of instability is a matter of opinion.  
The lower opacity of a  metal deficient atmosphere generally implies more compact stars. 
This could reduce the mass loss rate. 
However, we find (see Section 4) that, as a consequence of the second and third 
dredge up, the P-IMS become C-rich. 
The carbon excess alters the structure of the atmosphere which might favour 
 the formation of  
grains and increase the molecular blanketing,
so that the opacity and, in turn,  the mass loss rate could
grow.
Then, without an observational support, the actual mass loss rate that characterizes  
the AGB evolution of P-IMS is practically unknown. Note 
that the overall characteristics of the internal structure 
of an AGB star are generally dependent on two quantities: the core mass and the envelope mass
(see, for example, Iben \& Renzini 1983).
However while the envelope mass is larger than 1.5$-$2 $M_{\odot}$ 
the main parameter is the core mass. 
Thus, since the initial envelope mass of an intermediate mass star is rather large, 
the internal structure remains weakly dependent on the mass loss for a large fraction 
of the AGB phase. The AGB stellar models presented here have been computed without mass loss.
Our computations generally stopped 
before the onset of the mass loss-dominated AGB phase, even in the case of a particularly
strong mass loss (except for the last computed thermal pulses of the lowest mass model).

An even more complex situation concerns stellar convection in AGB stars. 
It has been widely discussed whether the classical
treatment of convection 
(both physical and numerical) is sufficient to describe
the second and third dredge up occurring during the AGB phase, or whether they require an extra-mixing  
whose physical nature has still to be clarified, acting below the convective border 
(Iben 1975, 1981; Becker \& Iben, 1979; Iben \& Renzini 1982, 1983; Hollowel \& Iben 1989; Lattanzio 1989;
Castellani et al. 1990; Castellani, Marconi \& Straniero 1998; Herwig et al. 1997;
Straniero et al 1997; Frost \& Lattanzio 1996: Langer et al. 1999; Mowlavi 1999; Herwig 2000). 
In summary, when the inner border of the convective envelope approaches the H depleted 
(He enriched) region, 
even a small perturbation, perhaps driven by a moderate mechanical overshoot,
may induce a substantial dredge up. 
In fact, as noted by Becker \& Iben (1979; see also Castellani et al. 1990 and
Frost \& Lattanzio 1996), if the convective envelope (H-rich) penetrates a region
 that is progressively more He enriched, 
a discontinuity of the radiative gradient forms 
at the interface between the stable and unstable layers. This occurs because the H-rich envelope has   a
significantly greater opacity
 than the He-rich layer immediately below it. In such a case,  if some mixing occurs below the 
inner border of the convective envelope,  
the local hydrogen abundance rises,  
the opacity (and the radiative gradient) grows and this layer becomes convectively unstable.
A similar situation is encountered at the outer edge of the convective
core during the central He-burning (Paczynsky 1970; Castellani, Giannone \& Renzini 1971); 
in this case the discontinuity of the radiative gradient 
is caused by the conversion of helium (low opacity) into carbon and oxygen (high opacity) at the base of 
the convective core.
Castellani et al. (1985) named 
this phenomenon {\it induced overshoot} (i.e. induced by the 
chemical discontinuity that forms at the boundary of a convective region)
 to be distinguished from the {\it mechanical overshoot} (i.e the mixing
caused by the convective elements that conserve a finite velocity beyond the unstable
region). Note that in the conditions discussed
above, the existence of a small (in most cases negligible) mechanical overshoot 
may trigger a substantial induced overshoot. 
Castellani, Marconi \& Straniero (1998) show that
the overall impact on the second dredge up of this phenomenon is generally small.
On the contrary, Frost \& Lattanzio (1996, see also Herwig et al 1997; Herwig 2000 
and Mowlavi 1999) found that the
inclusion of a moderate overshoot strongly increases the efficiency of the third dredge up.
Obviously a change in 
the strength of the third dredge up greatly affects the surface abundances of an AGB star.
Thus, spectroscopic observation of the 
present-day AGB stars may be used to check the reliability
of the various mixing algorithms and eventually to calibrate stellar models.
Unfortunately the same cannot be done with
the missing population III. Once again, due to the peculiarity of Z=0 models, the drescriptions 
obtained for more metal rich stars could be unreliable.
The basic set of models presented here has been obtained by limiting the
mixing within the borders defined by the Schwarzschild criterion. However we
have computed some additional models to investigate the influence of a possible extra-mixing
below the convective envelope on the AGB evolution of Z=0 stars.
Then, following to the hydrodynamic drescriptions discussed by Herwig 
et al. (1997), we have tentatively assumed that the convective motions
do not stop abruptly at the
base of the convective unstable envelope, but decrease exponentially below it. 
More precisely, we have used the following expression for the mixing velocity below 
the convective envelope:
\begin{equation}
v=v_{bce} exp(-D/\beta H_p)   
\end{equation}   
where $v_{bce}$ is the mixing velocity at the base of the convective unstable envelope (as obtained
by means of the mixing length theory), $D$ is the distance from the stability border, 
$H_p$ is the pressure scale height and $\beta$ is a free parameter,
which evidently determines the strength of the exponential decline. 
Herwig et al. (1997)
suggest $\beta=0.02$ (actually they use $f$ instead of $\beta$) for solar metallicity models.
We have used various values for $\beta$ ranging between 0.005 and 0.04 
(see Sections 4 and 5). 
Note that the extra-mixing provided by this formula is, in most cases,
negligible. In fact, since in the framework of the mixing length theory the mixing velocity is
proportional to the difference between the radiative and the adiabatic gradients, it
falls to zero at the boundary of a convective region, where the condition of marginal stability
(i.e. ${\nabla}_{rad}={\nabla}_{ad}$) is fulfilled.
However, when the convective envelope moves inward, down to the region where 
the H has been converted into He,
the discontinuity in the radiative gradient forms and $v_{bce}$ grows.   
In such a case, the extra-mixing provided by equation (5) 
smooths out the chemical profile and prevents the formation of the discontinuity
in the radiative gradient. In order to reduce possible numerical noises and save computational
 time, we do not apply equation (5) to the borders of other convective regions. 

\section {The central H and He-burning}

The overall characteristics of the central H and He burning phases of our models 
do not substantially  differ from the
ones already found in previous studies of very metal poor stars
(Ezer 1961, 1972; Ezer \& Cameron 1971; Eryurt-Ezer 1981; Chieffi  \& Tornamb\`e 1984;
Tornamb\`e \& Chieffi 1986; Cassisi and Castellani 1993). A more quantitative comparison with
the old models
by Ezer and coworkers reveals major differences probably due to differences 
 in the input physics. 
On the contrary, the models presented in the more recent papers 
are in good agreement with the present ones. This is no surprise because they were obtained
by means of old versions of our code (the FRANEC). 
Small residual differences can easily be attributed to slightly different input physics.

In Table 1 we report the duration (in Myr) of the most relevant evolutionary phases, namely 
the pre-main sequence (PMS), the central H-burning and
the central He-burning. Central temperatures versus central densities are plotted
in Figure 2, while the HR diagrams are shown in Figure 3. 
The temporal evolutions of the central convective region up to the end of the central He-burning 
are illustrated, for each star, in Figure 4.

As has been clear since the pioneering papers,
 the approach to the MS and the central H-burning phase are influenced by the
lack of CNO nuclei. In a more metal-rich IMS, in fact, 
the CNO cycle is the  main energy supplier
and since it has a very strong dependence on the temperature, the burning 
remains confined within the innermost part of the star and the convective core persists
until the fuel is almost exhausted in the centre.
In the case of these P-IMS, on the contrary, the nuclear energy might be produced only via the PP-chain (at least
 while the temperature remains below the threshold value for the activation of the $3\alpha$ 
 reactions) and since it has 
a weaker dependence on the temperature, the whole structure is necessarily much hotter and the active burning 
occurs in a more extended region around the centre.
The lack of a burning strongly concentrated in the centre leads to the following characteristics: 
a) the convective core is
much smaller than in more metal rich stars of similar mass and it even vanishes when the central 
 mass fraction of H is still 0.5, and b) in 
almost 80\% of the total mass of the star the He abundance increases. The
lack of an extended convective core clearly influences the stellar track in the  $T_C-{\rho}_C$ plane (see Fig. 2):
in particular,
since the PP-chain never slows the gravitational contraction effectively, the central temperature and density
continue to rise systematically until the $3\alpha$ reactions start up. The 
 abundance of carbon produced
in the centre as a function of time is shown in Figure 5.
When the $^{12}$C mass fraction is about $10^{-10}$ 
the H-burning switches to the CNO cycle. 
Then, the local luminosity immediately increases and
a new convective core appears (see Fig. 4).
The onset of the CNO burning induces the typical expansion of the central regions (see Fig. 2) and also
clearly marks the evolutionary tracks (see Fig. 3).

At the end of the H-burning the central temperature is so large ($\sim$ $10^8$ K) that
the He-burning immediately follows. These stars spend their central He-burning lifetime
at the blue side of the HR diagram, thus omitting the first dredge up episode.
For this reason, they enter the AGB phase with the original surface composition. 
As is usual at He ignition, a convective core develops, whose temporal evolution is shown in Figure 4. 
The H-burning shell, which
forms immediately outside the H exhausted core, is particularly hot. 
The evolution of the temperature and that of the location (in mass coordinates) of the mesh where
the energy generation rate of the H-burning  
is maximum, are shown in Figure 6 for the 7 M$_{\odot}$ model. Although the envelope is essentially 
 lacking metals,
the H-burning shell is mainly controlled by the CNO cycle. In fact, the temperature in the shell is
large enough to allow for some carbon production via
the 3$\alpha$ reactions. As a consequence, carbon is partially converted into nitrogen within the 
shell. The resulting internal profiles of hydrogen, carbon and nitrogen at the end
of the central He-burning are illustrated in Figure 7.

\section{The early AGB phase}

As is well known (see, for example, Becker \& Iben 1979) during the early-AGB the H and the 
He burning 
 are active in two 
separate shells. The energy generated by the He-shell induces an expansion 
  and a cooling of the outermost layers, so that a convective envelope, which penetrates 
 deeply within the star, develops. 
Though this is the first dredge up episode for these stars,  we will call it second dredge up, due to the  
similarity with the one found in the more metal rich models. 

The quantity (and the quality) of the matter dredged up to
the surface differs significantly from that of  more metal rich models and, as we will see below,
this derives in the formation of two families of AGB stars. 
Let us note first that the amount of He dredged up by these P-IMS is much 
larger than that found in 
the more metal rich AGB stars.  
 This is due to the fact that, since during most of the central hydrogen burning phase
the main energy source is the PP-chain, He is produced
in a wide region (in mass) of the stellar interior.
The second important thing to note is that, since the 3$\alpha$ reactions 
 are active up to the H-burning shell,
primary carbon is dredged up.  Some nitrogen also appears at the surface as a consequence of
the mixing 
of material processed by the H-burning shell. Note that this is the only case
in which the second dredge up alters the total abundance, by number, of the CNO group.
In the lower panel of Figure 8 we show the variations, during the early-AGB, 
of the locations of the H and He-burning shells (as identified by the points
 where the nuclear energy 
production is maximum)
as well as the extension of the convective envelope, 
for the 7 $M_{\odot}$ (no extra-mixing) model. 
The corresponding evolution of the surface abundances of C, N, O and He are shown in the
 upper panel. 
In Table 2 we have summarized some properties of our models at the
end of the early-AGB, namely (in columns 1 to 7): the stellar mass, the assumed value of
$\beta$ (see the definition of this parameter in section 2),
 the mass of the He core and the surface
mass fraction of $^{4}$He, $^{12}$C, $^{14}$N and $^{16}$O.
Note that the amount of CNO nuclei in the envelope of these stars 
 strongly depends on the initial mass.

Chieffi and Tornamb\'e (1984) (see also Fujimoto et al. 1984) have shown that a minimum
amount of CNO nuclei exists (roughly corresponding to a mass fraction of the order of $10^{-7}$) 
for which
the H-burning shell is fully sustained by the CNO cycle.
This means that the stars in which the CNO abundance 
in the envelope is raised above this threshold value by the second dredge up will have a more 
or less standard  
H-burning shell, while the masses in which the amount of CNO catalysts  
remain below this threshold value are forced to raise the temperature in the H-burning region  
up to that typical of the He ignition. Also, in the latter case, the CNO cycle dominates the
energy production in the H-burning
shell, but its efficiency is controlled by the amount of carbon locally produced via the 
$3\alpha$ reactions.
From the data reported in Table 2 it is evident that only for stellar masses larger than 
6 M$_{\odot}$ is 
this minimum amount of CNO  
attained in the envelope at the end of the early-AGB. In fact, in these models the 
expansion and the cooling induced by the He-burning shell  promptly  
allow a deep penetration by the convective envelope throughout 
the inter-shell region (see the lower panel of Fig. 8). For models having lower masses the inner
edge of the convective envelope penetrates only slightly (or not at all) the H-He discontinuity
and the resulting amount of CNO nuclei is definitely lower than the threshold value.      

The results shown in the first 4 rows of Table 2 refer to models obtained using the
Schwarzschild criterion to fix the boundaries of the regions mixed by convection. 
As discussed in section 2, we have also performed some tests to investigate the effects  
of a possible extra-mixing occurring below the border of the convective envelope. 
The results of these tests are given in the last 3 rows of Table 2.
They indicate that such an extra-mixing has only a negligible
effect on the efficiency of the second dredge up. A similar conclusion was previously
obtained by Castellani, Marconi \& Straniero (1998) for more metal-rich stars.

Before closing this section let us note that the 8 M$_\odot$ (Z=0) model  
ignites carbon off-centre during the early-AGB. 
We followed part of this C-burning up to the formation of an extended convective shell and 
 found that the minimum mass which is able to ignite carbon in a mildly degenerate core 
(usually called $M_{up}$)
is confined between the $7{\le}M_{up}/M_{\odot}{\le}8$ for a generation of stars of
 zero initial metallicity, which confirms the previous result obtained by Tornamb\`e \& Chieffi (1986) and
 Cassisi \& Castellani (1993).

\section{The advanced AGB evolution}

The early-AGB ends when the He shell, progressively approaching the H-He discontinuity, 
loses its efficiency and allows the overlying layers to contract and reheat. 
As a consequence,  
the H-burning shell reactivates and begins to accumulate fresh He on the underlying He core.
This is the beginning of the so called thermally pulsing AGB phase (henceforth TP-AGB).
When enough  He is accumulated, the $3\alpha$ reactions start
again at the base of the He-rich layer.
For the sake of completeness (for a more complete description of the TP-AGB stellar structures 
see the review of Iben \& Renzini 1983) let us briefly note that in stars 
with  {\it normal} metallicity  
the two shells (the H and the He ones) do not advance simultaneously in mass but, on the contrary,
 are active alternately. The He ignition is characterized by a rather strong 
thermonuclear runaway. The conditions of this He shell flash are somewhat different 
from those that  lead to the He flash  
in a low mass star at the tip of the Red Giant Branch. The thermonuclear runaway, in the latter case, is  induced by the strong degeneracy of the 
electron component of the stellar plasma, while in TP-AGB
stars the He-rich layer is largely non degenerate. If matter is non degenerate,
the tendency for temperature to increase, due to the local release of thermonuclear energy,
is normally counterbalanced by an expansion caused by the temperature dependence of the pressure.  
In most cases this pressure response
 rapidly quenches the heating 
and a quiescent (self-regulated) burning takes place. 
However, in TP-AGB models, the He ignition occurs in particular conditions
(see Schwarzschild \& Harm 1965) 
 and the local temperature may rise for a certain time (how much mainly depends on
the core mass) before the
pressure response becomes effective. As the $3\alpha$ reaction rate is strongly
dependent on the temperature, the energy production rate of the He-burning shell
rapidly increases up to $10^6-10^8 L_{\odot}$.
The strength of the He flash depends, among other things,
on the local values of temperature and density 
of the He-rich layer  at the moment of the fuel ignition. For a given core mass,
lower temperatures and larger densities imply higher He flash luminosities. 
It is important to note that these conditions are influenced by the rate 
at which the
He-rich layer is accreted and, hence, by the properties of the H-burning shell.
It is therefore clear that the amount 
of CNO catalysts in the envelope is a fundamental quantity that affects
many properties of the TP-AGB phase, such as: the duration of the inter-pulse
($\Delta t_{ip}$), the maximum $3\alpha$ luminosity and the growth rate of the C-O core mass. 
For all these reasons one may expect a particular behaviour of the Z=0 TP-AGB stars.

As already mentioned, Chieffi \& Tornamb\'e (1984) did not find thermal instabilities
 during the AGB of 
a 5 M$_{\odot}$ (Z=0) stellar model.
In fact they showed that, owing to the lack of CNO nuclei, the temperature in the H-burning shell
 is so high that
the 3$\alpha$ reactions are active. Since the temperature at the base of the He-rich 
layer must be
even higher, the two shells must be simultaneously active. In a companion paper
Fujimoto et al. (1984) showed that under this condition, the stability of the He-burning shell depends
on the core mass: those stars having a core mass larger than a critical value 
($\sim$ 0.73 M$_{\odot}$) develop a stable He-burning shell, whereas for lower core masses thermal 
pulses occur. This prediction was confirmed by Chieffi \& Tornamb\'e: in their 5 M$_{\odot}$ model 
the two shells did advance (in mass) by consuming fuel at the same rate 
(at least within the phase that it was possible to compute at that time),
so that a stable steady state was achieved during the AGB. 

Concerning  our set of P-IMS, let us discuss separately,
for the sake of clarity, the more massive models, 
in which the second dredge-up is able to raise the CNO mass fraction in the envelope above 
the threshold value for a self-sustained CNO burning (i.e. $\sim$ $10^{-7}$), 
and the less massive ones.
It is important to note that all the present models develop 
a core mass, at the end of the early-AGB, larger than the critical value obtained 
by Fujimoto et al. (1984).

\subsection{6$\le$M/M$_{\odot}$$<$M$_{up}$}

In these models the
CNO cycle in the H-burning shell reaches its full efficiency at a temperature lower than
 the value  needed
to activate the 3$\alpha$ reactions.
In this case the development of the thermally pulsing phase proceeds qualitatively as in the 
more metal rich 
AGB stars. The luminosity contributions of both the H and the He-burning shells, 
in the case of the
7 $M_{\odot}$ model and for two different values of the $\beta$ parameter (see Section 2 and below),
are shown in Figures 9 and 10.
In Figures 11 and 12 we show  the evolution of the corresponding locations
(in mass coordinates) of the two shells as well as the extension of
 the convective envelope. Moreover some properties of 
these thermally pulsing models are listed in Tables 2 and 3, namely (in columns 1 to 8):
the position of the H-burning shell at the onset of the thermal pulse (exactly 
when the He-burning luminosity
becomes larger than the H-burning luminosity),
the duration of the inter-pulse,
the peak luminosity of the He-flash, the temperature of the H-burning shell 
(exactly at the point where the nuclear energy production is maximum) at He re-ignition,
 the density of the He-burning shell at He re-ignition and 
the surface mass fractions of $^{12}$C, $^{14}$N and $^{16}$O.
Firstly we note that, 
owing to the lower amount of CNO nuclei in the envelope,
the temperature of the H-burning shell is higher than that typically found in more metal-rich
models (see, for example, Straniero et al. 2000). For example, in a 7 $M_{\odot}$ model (Z=0.02) we found,
at the beginning of the $10^{th}$ TP, that the temperature of the H-burning shell is
$\log T_H=7.93$, while in the present Z=0, even though the core mass is slightly lower,
we find $\log T_H=8.10$.  
As a consequence, the temperature of the He accumulated by the H-burning shell is higher 
and, in turn, a smaller density
is required to start the 3$\alpha$ reactions. This occurrence becomes more evident if
the properties of the He shell at the onset of a TP are compared to the corresponding 
ones in a model
whose envelope has been enriched with fresh CNO catalyst. 
 Figure 13 shows the evolution of the temperature, the density and the pressure of the He
shell in the period between the $9^{th}$ and the $12^{th}$ TP of the
7 $M_{\odot}$ and $\beta=0.005$ model. In this period the surface CNO mass fraction
is approximately constant ($\sim$2.2 $10^{-6}$). The same quantities are given 
in Figure 14, but for the last computed TPs. In the latter case the mass fraction of CNO in the envelope
 grows from about 
6.4 $10^{-5}$ to 1.33 $10^{-4}$. Then, the larger the CNO abundance in the envelope the lower
the temperature and the larger the density
attained by the He-rich layer before the beginning of a TP. Note that in this case 
the duration of the inter-pulse is longer, 
because more time is needed to reach the conditions for the He ignition. 
The greater density of the He-rich material at the moment of the re-ignition
clearly implies a stronger thermal pulse.
This is evident from Figures 9 and 10 (see also the third column in Tables 3 and 4).

In spite of the lower intensity of the thermonuclear runaway, in the present Z=0 models, too,  
the He flashes 
induce the formation of a convective region which extends, after a few TPs, 
over the whole inter-shell region. In such a way, the products of the He-burning
(essentially carbon)
are distributed throughout the H-exhausted region.
After each pulse, when the quiescent He-burning takes place at the base of the He-rich layer,
the inner border of the convective envelope moves inward and closely approaches
the H-He discontinuity. For example, at the $10^{th}$ TP of the 7 M$_{\odot}$ models,
the inner  
border of the convective envelope penetrates the H-burning shell reaching a layer which is only
$\sim$$10^{-5}$ $M_{\odot}$ away from the C-enhanced region.
Note that this situation has also been found in the case of a mixing strictly confined within the
boundary defined by the Schwarzschild criterion.   
Becker and Iben (1979) were 
first to note that this is a very promising case, in which an overshoot 
may be activated owing to the {\it  
"finite and positive value of ${\nabla}_{rad}-{\nabla}_{ad}$ at the base of the formal convective
zone"}. They found that this happens when the amount of He-rich material that remains below 
the convective envelope is less than about 0.1 $M_{\odot}$. Such a condition is very soon fulfilled 
(after about 4 or 5 TPs) in our Z=0 models that have a  mass larger 
than 6 $M_{\odot}$.
The resulting {\it induced overshoot} may be of great relevance for the possible dredge up
of carbon-rich material. This carbon dredge up, which is commonly named 
the third dredge up (henceforth TDU), has been invoked to explain 
the formation of the present generation of carbon stars (Iben 1975) and it is strongly supported 
by the observational evidence that there exists an evolutionary sequence of the 
various AGB components (from  M to S and C stars) characterized by a progressive increase in  
the C/O ratio. As explained in Section 2 we have activated this induced overshoot 
by artificially adding
a small extra-mixing below the base of the convective envelope. In order to check the 
dependence of the resulting dredge up from the assumed strength of the extra-mixing 
we have computed various models by changing the free parameter $\beta$ in 
equation (5). 
In all the cases in which an extra-mixing below the formal convective envelope has
been included, a deep carbon dredge up is obtained. Note the two different regimes
that characterize the
TP-AGB evolution before and after the first C-dredge up episode (see Figures 9 to 12 and 
Tables 3 and 4). For
few TPs, the C-dredge up induces a rapid H re-ignition (this is evident in the upper panels
of Figures 9 and 10). 
For $\beta \ge 0.02$ the first
 H thermonuclear runaway is particularly
strong (the peak luminosity is larger than 5 $10^6$ $L_{\odot}$).
 To follow this phase our adaptive mesh algorithm
puts more than 5000 mesh points in the convective envelope, 3000 just in the 
innermost 0.1 $M_{\odot}$. In this case the mixing algorithm described
in Section 2 consumes a huge amount of computer time (about 3 days to compute 
one dredge up episode on a 
700 MHz workstation). For this reason we have followed only two C-dredge up episodes
in the $\beta = 0.02$ case and just one in the ${\beta}=0.04$ case. Note that the total 
amount of CNO dredged up
in the first episodes of C-ingestion in the $\beta = 0.02$ case (about 5 $10^{-4}$ $M_{\odot}$)
is equivalent to that
cumulatively
ingested during five dredge up episodes in the $\beta=0.005$ case. It is clear 
that, owing to the increase in the CNO in the envelope, the temperature of the H-burning
shell and the strength of this H flash are progressively reduced (see column 4 of Tables 3 and 4
and Figures 9 and 10)
until this peculiarity of the Z=0 TP-AGB models also disappears. 
Let us finally note that Herwig (2000) did not find a clear dependence of the dredge up efficiency 
on the extra-mixing efficiency. Perhaps this discrepancy may be due to the different 
chemical composition of the Herwig models (namely Z=0.02). 
Nevertheless his tests were made by changing the parameter that control
the extra-mixing for just one TP. We suspect that a more evident variation of the dredge up
 should be found if the
AGB evolution is followed for several TPs (as we actually do in the present paper) 
under different assumption for the extra-mixing efficiency. In any case,
the great relevance of the dredge up on our capability of predict the AGB properties 
demands a deeper analysis of this question which is evidently beyond the scope
 of the present paper.

\subsection{4$\le$M/M$_{\odot}$$<$6}

The scenario described above holds for the more massive P-IMS, for which 
the CNO abundance
in the envelope is raised by the second dredge up above the threshold value
which allows the full efficiency of the H-burning shell
without the requirement of additional CNO catalysts coming from a local production of carbon.
This condition implies the formation of fairly normal AGB stars.
On the contrary, the behaviour of the less massive Z=0 stars is rather peculiar. 

Figure 15 shows,
for a 4 M$_{\odot}$ ($\beta = 0.01$) model, the H and He luminosities as a function of the time, while Figure 16
shows the locations (in mass coordinates) of the He and the H-burning shells  and the
extension of the convective envelope  as a function of time. Table 5 lists 
some properties of the same model.

The first thing worth noting is that, though the CNO mass fraction in the envelope
is lower than $10^{-7}$ and though the core mass 
is larger than the critical upper limit found by Fujimoto et al. (1984), 
these  AGB models experience thermal instabilities. 
The H-burning is indeed regulated by the 
carbon locally produced via the $3\alpha$ reactions, but   
the coupling of the H and the He-burning is so weak that they do not advance
(in mass) at the same rate. 
As a consequence, small amplitude thermal pulses take place. 
 Owing to the weakness of these He-shell flashes, the inner border of the convective
 envelope does not penetrate the He-rich layer so that the induced overshoot
 is probably not activated. 

The second, very important occurrence is that since the first thermal pulse 
the H shell becomes unstable and a convective zone develops 
(hereinafter HCE: Hydrogen Convective Episode), 
whose maximum extension is attained just after the usual He convective episode (HeCE)  
starts to shrink and just before the convective envelope reaches its maximum 
inward penetration. 
Let us stress that these new convective episodes are a systematic property of these
Z=0 AGB models with masses ranging between 4 and 6 M$_{\odot}$.
The base of this HCE closely approaches the border of the carbon-enriched
zone left by the HeCE in the inter-shell. This condition is similar to the 
one already encountered when the convective envelope penetrates the H shell
(see previous subsection). In fact, owing to the greater opacity
of the H-rich material, a discontinuity of the radiative gradient forms at the
interface
between the base of the HCE and the He-rich layer below. This 
condition is favourable to the induced overshoot that, in our opinion,
 would easily extend the base of the HCE 
down to the C-enhanced layer.      
In such a case, the carbon
ingestion would induce a sudden thermonuclear runaway of the H-burning shell. 

The consequences of this mixing event are illustrated in Figures 
16 and 17. 
In this particular case the two
shells move simultaneously outward for some time, but, 
when the innermost burning front gets closer to the
H-He discontinuity, its rate relaxes. From now on, weak thermal pulses take place, 
characterized by the 
two convective episodes described above: the standard HeCE and the unusual HCE.
During the $5^{th}$ TP the HCE extends inward and overlaps
the region previously mixed 
by the He convective shell. This situation is illustrated in Figure 17. 
Then carbon is dredged up and immediately a H 
flash occurs. The energy released by this second (H) flash makes an important contribution to 
the expansion of the outermost layers started by the first (He) flash. 
Later on, during the post-flash period, the convective envelope penetrates 
the H-burning region. 
Thus, the surface abundances of C and N are significantly enhanced 
(the CNO mass fraction in the envelope becomes $\sim 4 \cdot 10^{-6}$). 
This is an irreversible 
phenomenon since the amount of C dredged up is now so large that the star behaves, 
from now on, 
like the more massive ones.
In particular after a few more TPs, the {\it standard} TDU takes place.    

Note that in a model computed with a mixing
strictly confined within the boundaries formally defined by the Schwarzschild criterion,
the base of the HCE comes very close to the ex-convective shell generated by the 
thermal pulse and the occurrence of the carbon dredge up is arbitrarily dependent
on the accuracy of the computation. Our experiments show that by 
changing the number of time steps and/or 
the spatial resolution, the HCE may or may not penetrate the C-rich region below it.
 However, once a small extra-mixing 
is included (even just by putting $\beta=0.001$ in equation (5)) 
this uncertainty disappears and the carbon dredge up at the base of the HCE 
is systematically obtained after a few TPs.
In the particular case illustrated in figures 15, 16 and 17 a first close approach of the 
HCE to the C-rich inter-shell region 
has been obtained after just 3 TPs (see column 6 of table 5), but only the fifth one 
is capable to dredge up enough C to turn off the AGB evolution 
of this model.  
 In our opinion the case in which the HCE does not enter 
the carbon rich layers is unrealistic since it comes, in any case, so close to the 
C-rich region that any
small instability would lead to the 
irreversible and definitive mixing that
 brings the 
AGB properties of these models towards those of the more massive ones.

\section{Conclusions}
 
The occurrence of thermal pulses in primordial AGB stars with masses ranging between 
4 and 8  M$_\odot$ implies a major revision of the early nucleosynthesis scenario.
We have shown that models with M$\ge$6 $M_{\odot}$ develop a 
deep convective envelope that penetrates the H shell and approaches the region
previously enriched with the carbon produced during the He shell flash.
Moreover we have demonstrated that 
in this condition 
even a small extra-mixing occurring below the base of the convective envelope induces a significant
carbon dredge up. We note that the abundances measured in the present generation of
AGB stars demand the occurrence of the third dredge up
(see, for example, the review of Iben \& Renzini 1983 and references therein).     
Then, the fresh
carbon engulfed in the envelope is partially converted into nitrogen by the
 CNO burning
 occurring at the base of the
convective region. Thus these AGB stars develop C and N rich envelopes. 
     
Our models with masses ranging between 4 and 6 $M_{\odot}$ start the TP-AGB phase with weak He shell
flashes so that the base of the convective envelope
remains far from the H-burning shell. 
We show, however, that immediately after a thermal pulse,
the H-burning shell becomes convectively unstable and the inner
border of this convective region (HCE) closely approaches the C-enriched zone.
Also, in this case, 
 we believe an induced overshoot is likely to occur. In our models, which include 
a small additional mixing at the base of the HCE, a thermonuclear 
runaway is activated by the carbon ingested by the HCE. The resulting expansion of the outermost
layers leads to the deep penetration of the convective envelope that overlaps the whole region
previously mixed by the HCE. Then the abundance of CNO in the envelope rises enough to allow
 a 
{\it normal} AGB evolution from now on. 
In particular, after a few TPs, the TDU may further increase the carbon abundance in the envelope 
so that, once again,
it is partially converted into nitrogen by the CNO burning occurring at the base of the 
external convective region.  

The precise evaluation of the amount of nitrogen and carbon produced by these primordial AGB
stars
depends on the
value of the parameter used 
to describe the strength of the extra-mixing ($\beta$ in our equation (5)).
 The uncertainty concerning the 
mass loss rate is a further (important) limit to our comprehension of the nucleosynthesis
contribution from these stars. 
Any attempt to calibrate these free parameters encounters 
 the problem of the scarcity of observable constraints for population III stars. Note, however, that
though a quantitative derivation of the nucleosynthesis products requires a precise 
determination of the efficiency of mass loss and convective mixing,
the new evolutionary scenario for P-IMS emerging from our models is not substantially
 affected by the above uncertainty.  

The imprint of this nucleosynthesis could be sought in the more metal deficient
stars presently observed in our Galaxy and in the high redshift Lyman $\alpha$ systems.
Observations show solar and relatively flat [C/Fe] and [N/Fe] 
ratios with metallicity which indicates a primary
origin of both elements (see, for example, McWilliam 1997 and references therein).
 The nucleosynthesis of massive stars is sufficient to explain the primary carbon
observed in metal poor halo dwarfs, but it cannot account for the primary nitrogen
 (Timmes, Woosley \& Weaver 1995).
In a companion paper (Abia et al. 2001) we investigate the 
implications of the new models for population III stars  
on the chemical evolution of the early galaxy.
We show that the P-IMS models presented here, coupled with our new models
for massive Z=0 stars (Limongi, Straniero \& Chieffi 2001) allow us
 to reproduce the primary
behaviour of both C and N down to the lowest observed metallicity.
For example, by using the P-IMS computed assuming $\beta=0.01$ 
we obtain -0.3$\le$[N/Fe]$\le$+1.0, 
the precise value depending on the assumed initial mass function and on the iron amount
ejected by Z=0 SNII (see also Dom\'\i nguez et al. 2000).
 
Recently, it has been pointed out that a considerable fraction (20-25$\%$) of the most metal poor
stars are carbon-rich, [C/Fe]$\ge$1  (Rossy, Beers \& Sneden 1999). In most cases these extremely metal
 poor carbon stars
are also nitrogen-enhanced
(Norris, Ryan \& Beers 1997; Bonifacio et al. 1998; Hill et al. 2000). Fujimoto, Ikeda \& Iben (2000) suggest that some of these carbon stars may be the 
product of mass transfer from an AGB companion of mass lower than 3 $M_{\odot}$ in close binary systems.
Our computations 
show that in the case of an intermediate mass companion, too, the surface composition
of the secondary star after the mass transfer episode should be CN enhanced.

Another interesting product of the nucleosynthesis occurring in these P-IMS models is 
Li. When the temperature at the base of the convective envelope is higher than 
$20-30 \cdot 10^6$ K,  beryllium
is efficiently produced through the $^3$He($^4$He,$\gamma$)$^7$Be reaction.  
Then, $^7$Li is synthesized via the electron capture of $^7$Be, whose terrestrial half life is
about 53 days.  If this electron capture occurs at the base of the envelope, the resulting
Li would be immediately destroyed as
a consequence of the $^7$Li(p,$\alpha$)$^4$He reaction. However the convective motion might be 
so fast that most of the beryllium produced at the base of the envelope is redistributed 
in the outermost layers of the star, 
where the temperature is lower and Li is preserved. 
This is the well known Cameron-Fowler mechanism (Cameron 1955).
In our models of P-IMS such a process is particularly efficient during the TP-AGB phase so that
the mass fraction of Li in the envelope increases to a value
comparable with the one expected from the big bang nucleosynthesis (see, for example,  Walker et al. 1991).  
 
Let us finally mention an interesting nucleosynthesis channel that may be activated
in zero metal AGB stars by the 
production of free neutrons. As is well known (see, for example, Busso, Gallino \& Wasserburg 1999), 
the main component of the cosmic s-elements has been synthesized 
by neutron captures on iron seeds occurring in TP-AGB stars. Being deprived of iron,
P-IMS cannot contribute to this nucleosynthesis. However,   
if  the two major neutron sources,
namely the 
$^{22}$Ne($\alpha$,n)$^{25}$Mg and the  $^{13}$C($\alpha$,n)$^{16}$O reactions, 
are activated they might contribute to the synthesis of lighter elements. 
We found that the $^{14}$N left by the advancing H-burning shell 
is fully converted into $^{22}$Ne during the TP. Obviously, as this nitrogen
 depends on the amount
of CNO in the envelope, the amount of $^{22}$Ne in the convective region generated by the 
He flash is rather small at the beginning of the TP phase. However, after some dredge up episodes 
this value increases significantly.
As an example, in the last computed TP of the 4 $M_{\odot}$ ($\beta=0.01$), the mass fraction of
$^{22}$Ne in the inter-shell region is about $5 \cdot 10^{-3}$, which is about half the value 
typically found in  AGB models of solar chemical composition. 
In our Z=0 models the maximum temperature at the base of the convective zone generated
 by the He shell flash 
ranges between 310 and $350 \cdot 10^6$ K  so that the $^{22}$Ne($\alpha$,n)$^{25}$Mg
 is definitely involved.
Since the amount of $^{22}$Ne in the inter-shell increases with time and since the
temperature at the bottom of the convective shell also increases, the neutron production and the 
related nucleosynthesis would be particularly  
efficient in the final part of the TP-AGB phase of these P-IMS. 
For these reasons many more models are needed to calculate this neutron capture 
synthesis. Nevertheless the release of free neutrons 
(at least from the $^{22}$Ne source) in an environment lacking iron seeds should activate
a particular nucleosynthesis channel whose product could be used to trace the pollution caused
by the missing  population III in the primordial galactic material.

\acknowledgments
This work has been partially supported by the MURST Italian grant Cofin2000,
by the MEC Spanish grant PB96-1428, by the Andalusian grant FQM-108 
and it is part of the ITALY-SPAIN
integrated action (MURST-MEC agreement) HI1998-0095.  
A. Chieffi thanks the Astronomical Observatory of Rome and its Director, Prof. Roberto Buonanno,
for the generous hospitality at Monteporzio Catone. I. Dom\'\i nguez thanks
everybody from the {\it Osservatorio Astronomico di Collurania} (Teramo, Italy) for their
hospitality.

\newpage

\centerline{\bf FIGURE CAPTIONS}
\vspace{1cm}

\figcaption[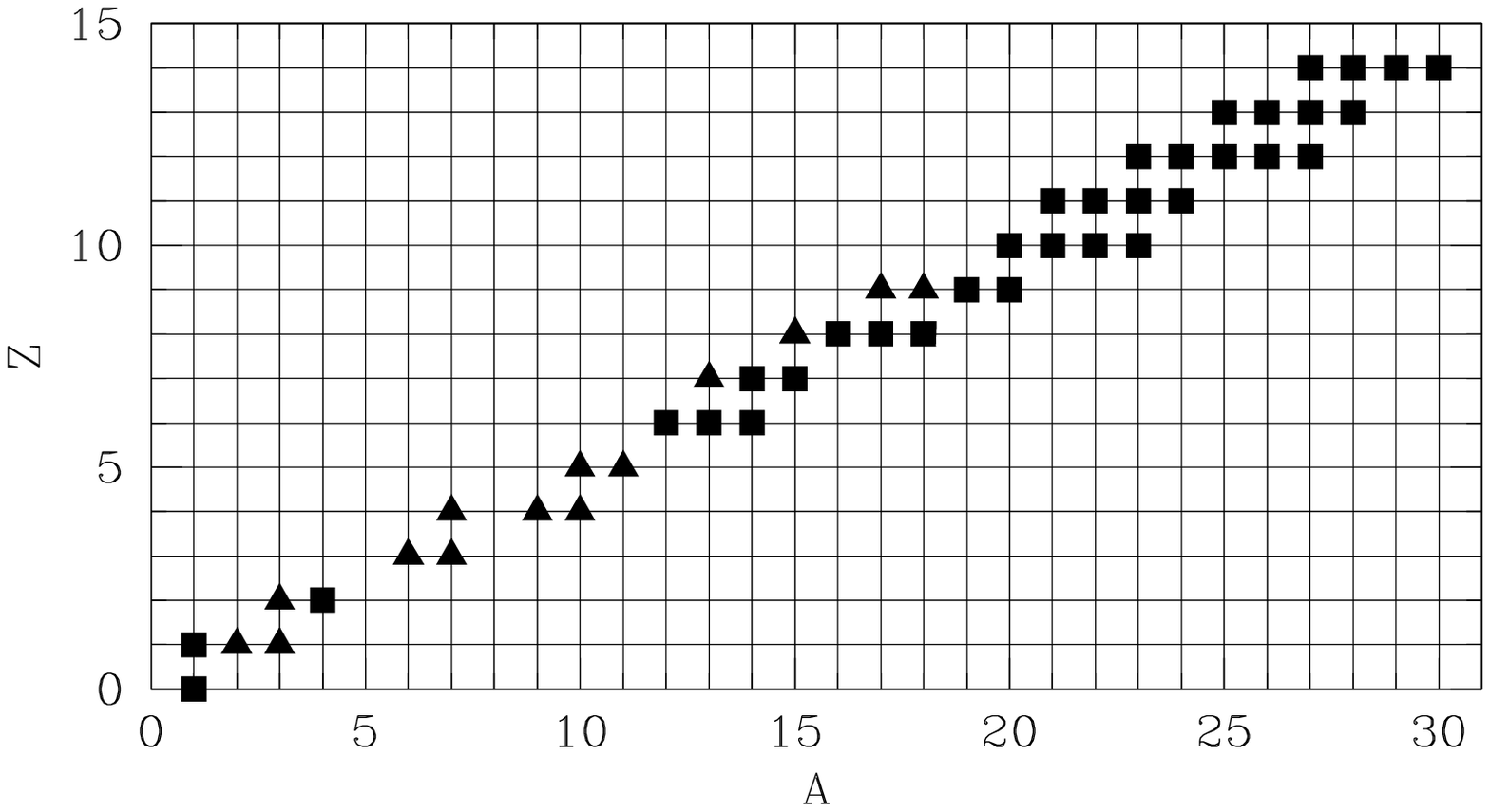]{Isotopes included in the nuclear network; those identified by     
 triangles have only been included during the H-burning. 
\label{f1int}}

\figcaption[fig_2.eps]{Evolution of the central temperature as a function of the
  central density for the computed models.
\label{f2int}}

\figcaption[fig_3.eps]{Evolutionary tracks in the HR diagram of the computed models. 
\label{f3int}}

\figcaption[fig_4.eps]{Evolution with time of the central convective regions up 
 to the end of the He burning for the computed models:  H-burning via 
 the PP-chains (first episode),  H-burning via CNO cycle (second episode) 
 and  He-burning (third episode). 
\label{f4int}}

\figcaption[fig_5.eps]{Evolution with time of the central C abundance (mass fraction) 
 for the computed models.  
\label{f5int}}

\figcaption[fig_6.eps]{Evolution with time of the H-burning shell during the central He-burning phase
for the 7 M$_{\odot}$ model. 
Lower panel: the position (in mass) 
where the nuclear energy generated by the H-burning shell is maximum.
 Upper panel: the temperature at the same point. 
\label{f6int}}

\figcaption[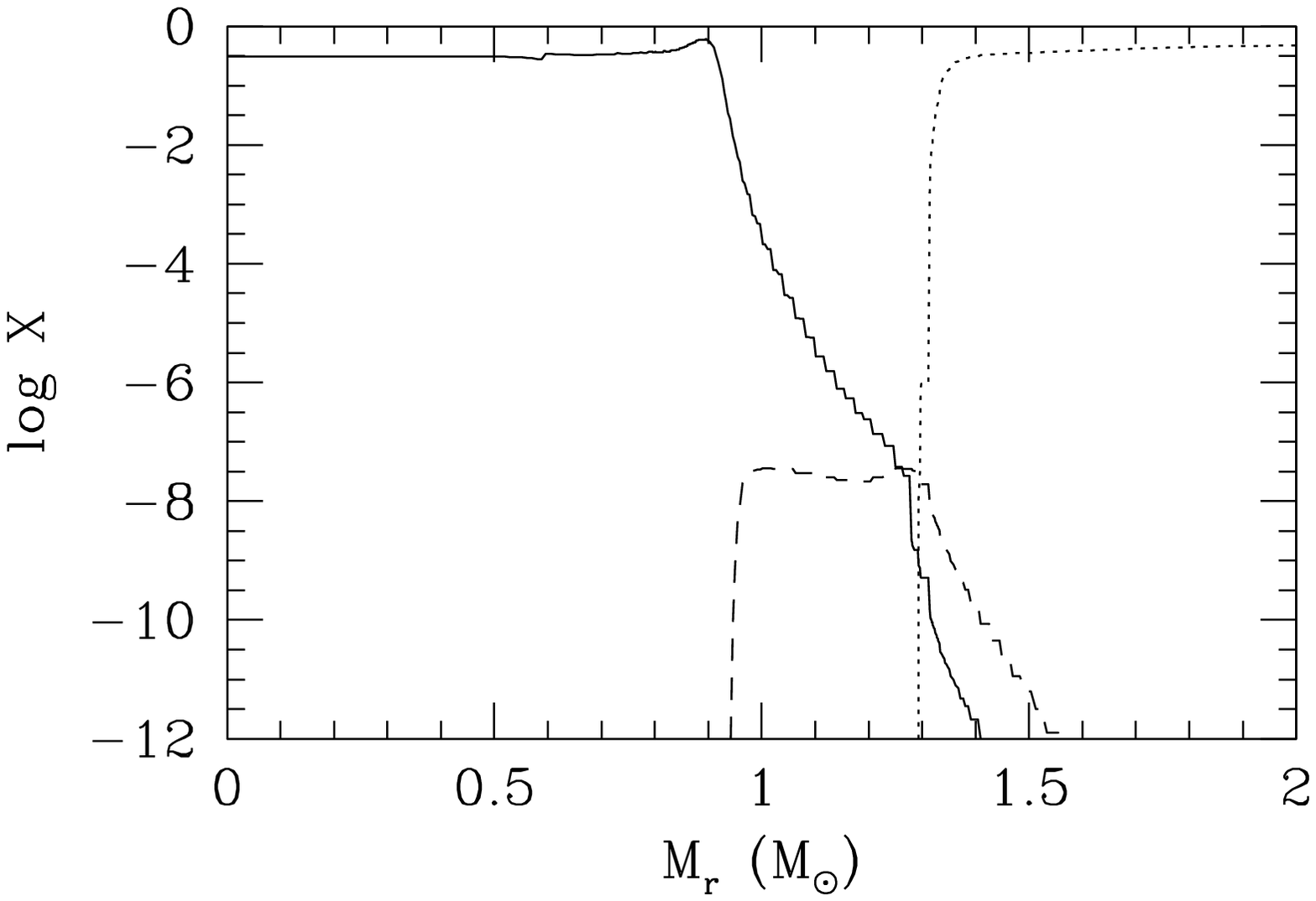]{C (solid line), N (dashed line) and H (dotted line) mass fraction profiles 
 of the 7 M$_{\odot}$ model at the end of the central He burning.
\label{f7int}}

\figcaption[fig_8.eps]{The early-AGB phase  
 of the 7 M$_{\odot}$ model (no extra-mixing). Upper panel: 
 evolution with time of the surface abundance (mass fraction) 
 of C (long-dashed line), N (short-dashed line), O (dotted line) and 
  He (solid line). Note that the He scale reads on the right Y axis.  
 Lower panel:  evolution with time of the location (in mass coordinates) of the 
 H and He-burning shells. The dashed area shows the extension of the convective envelope.
\label{f8int}}

\figcaption[fig_9.eps]{Evolution with time of the H and He-burning luminosities
 during the TP-AGB phase in the case of the 7 $M_{\odot}$ ($\beta = 0.005$) model. The initial time corresponds to the deepest penetration of the convective envelope after the $2^{nd}$ dredge up. 
\label{f9int}}

\figcaption[fig_10.eps]{Evolution with time of the H and He-burning luminosities
 during the TP-AGB phase in the case of the 7 $M_{\odot}$ ($\beta = 0.01$) model. The initial time corresponds to the deepest penetration of the convective envelope after the $2^{nd}$ dredge up.
\label{f10int}}

\figcaption[fig_11.eps]{Evolution with time of the location (in mass coordinates) of the  
 H and the He-burning shells of the same models as in Figure 9. 
 The dashed area shows the extension of the 
 convective envelope.
\label{f11int}}

\figcaption[fig_12.eps]{Evolution with time of the location (in mass coordinates) of the  
 H and the He-burning shells of the same models as in Figure 10. 
 The dashed area shows the extension of the 
 convective envelope.
\label{f12int}}

\figcaption[fig_13.eps]{Properties of the He-burning shell between the $9^{th}$ and the 
$12^{th}$ TP of the 7 $M_{\odot}$ ($\beta = 0.005$) model: density (upper panel). temperature
(central panel) and pressure (lower panel). The initial time corresponds to the deepest penetration of the convective envelope after the $2^{nd}$ dredge up.
\label{f13int}}

\figcaption[fig_14.eps]{Properties of the He-burning shell between the $17^{th}$ and the 
$20^{th}$ TP of the 7 $M_{\odot}$ ($\beta = 0.005$) model: density (upper panel), temperature
(central panel) and pressure (lower panel). The initial time corresponds to the deepest penetration of the convective envelope after the $2^{nd}$ dredge up.
\label{f14int}}

\figcaption[fig_15.eps]{Evolution with time of the H and He-burning luminosities
 during the TP-AGB phase in the case of the 4 $M_{\odot}$ ($\beta = 0.01$) model. The initial time corresponds to the deepest penetration of the convective envelope after the $2^{nd}$ dredge up.
\label{f15int}}

\figcaption[fig_16.eps]{Evolution with time of the locations (in mass coordinates) of the  
 H and the He-burning shells of the same models as in Figure 15. 
 The dashed area shows the extension of the 
 convective envelope. 
\label{f16int}}

\figcaption[fig_17.eps]{Evolution with time of the convective episodes (dashed regions) 
 during (and immediately after) the 
 $4^{th}$ and $5^{th}$ TP of the same models as in Figure 15.  The two (approximately) horizontal
 lines represent the location of the H and He-burning shells. A first convective episode (HeCE)
 develops in the inter-shell region as a consequence of the TP. A second one starts later
 at the base of the H-rich envelope (HCE). During the $5^{th}$ TP the innermost part of the HCE  
 overlaps  
 the region previously mixed by the HeCE; this induces two H flashes that produce the two
 evident tongues of the outer border of the HCE. 
 Later on, the convective envelope moves inward and a dredge up of C and N occurs.
\label{f17int}}

\end{document}